\documentstyle[aps,prb,multicol]{revtex}

\begin{document}

\newcommand{\uprule}{\end{multicols}
\noindent \vrule width3.375in height.2pt depth.2pt 
\vrule height.5em depth.2pt \hfill \widetext }
\newcommand{\downrule}{\indent \hfill \vrule depth.5em height0pt 
\vrule width3.375in height.2pt depth.2pt 
\begin{multicols}{2} \narrowtext}	
	
\include{psfig}

\draft \title{\bf Anisotropic renormalized fluctuations in the 
microwave resistivity in YBa$_2$Cu$_3$O$_{7-\delta}$}

\author{D.Neri$^{(1,2)}$, E.Silva$^{(1)}$, S.Sarti$^{(3)}$, 
R.Marcon$^{(1)}$, M.Giura$^{(3)}$, R.Fastampa$^{(3)}$ and 
N.Sparvieri$^{(4)}$ }

\address{ $^{(1)}$Dipartimento di Fisica ``E.Amaldi'' and Unit\`{a}  INFM,\\
Universit\`{a} ``Roma Tre'', Via della Vasca Navale 84, 00146 Roma, 
Italy\\
$^{(2)}$Dipartimento di Ingegneria Elettronica,\\ Universit\`{a} ``Roma Tre'', Via 
della Vasca Navale 84, 00146 Roma, Italy\\
$^{(3)}$Dipartimento di Fisica, Unit\`{a}  INFM,\\ Universit\`{a} ``La Sapienza'',
P.le Aldo Moro 2, 00185 Roma, Italy\\
$^{(4)}$Alenia Direzione Ricerche, Via Tiburtina km 12,400, 00131 
Roma, Italy}

\date{accepted for publication, Physical Review B February 6, 1998; revised June 25, 1998} \maketitle

\begin{abstract}
We discuss the excess conductivity above $T_c$ due to renormalized 
order-parameter fluctuations in YBa$_2$Cu$_3$O$_{7-\delta}$ (YBCO) at
microwave frequencies. 
We calculate the effects of the uniaxial anisotropy on the 
renormalized fluctuations in the Hartree approximation, extending the 
isotropic theory developed by Dorsey [{\em
Phys. Rev. B} {\bf 43}, 7575 (1991)].
Measurements of the real
part of the microwave resistivity at 24 and 48 GHz and of the dc 
resistivity are performed on different YBCO films. The
onset of the superconducting transition and the deviation from the 
linear temperature behavior above $T_c$ can be
fully accounted for by the extended theory. According to the 
theoretical calculation here presented, a departure
from gaussian toward renormalized fluctuations is observed. Very 
consistent values of the fundamental parameters
(critical temperature, coherence lenghts, penetration depth) of the 
superconducting state are obtained.
\end{abstract}


\begin{multicols}{2}

\section{Introduction}
\label{intro}

The analysis of fluctuations-induced excess conductivity has 
stimulated in the past years a considerable amount of
work. Theoretical investigations of the dc as well as the 
finite-frequency conductivity  dates from the 1960's,
\cite{skoc,asla,schmidt,klemm,maki} and development of this topic 
proceeded until the discovery of high temperature
superconductors (HTSC's). High critical temperatures and short 
coherence lengths conspire to the giant
enhancement of thermodynamical fluctuations in HTCs. Due to their 
layered structure and to the consequent
anisotropy in the superconducting state, the effects of thermal 
fluctuations are further enhanced. As a result, these
materials can be viewed as ideal systems to experimentally verify the 
theories for the excess conductivity. Among
the high-temperature compounds with moderate but significant 
anisotropy, YBCO is the most
studied on this aspect,
\cite{ybco,holm,hopf,booth,naki,kamal,andreone,anlage,ike,gauzzi} 
and we will restrict our
discussion and measurements to this compound only.\\
Despite  the extended experimental investigation, there is a 
considerable debate on the appropriate model for
the fluctuation conductivity in zero magnetic field:
\cite{holm,hopf,gauzzi,cimberle,varlamov} most of the existing
analysis of the effects of order-parameter fluctuations on the dc 
conductivity have been performed in terms of
Aslamazov-Larkin (AL) isotropic three-dimensional (3D) fluctuations 
\cite{asla} for temperatures close to $T_c$,
while a crossover to 2D fluctuations has been claimed at temperatures 
substantially higher than $T_c$,\cite{holm}
according to the interpretation of the data in terms of 
Lawrence-Doniach (LD) model.\cite{ld} Not too close to $T_c$,
data on YBCO have been shown to be compatible with an AL 
interpretation supplemented with a Maki-Thompson
(MT) term.\cite{maki,holm} However this framework has been seriously 
questioned by Hopfengartner \textit{et al}:
\cite{hopf} by extending the AL theory to an anisotropic 
superconductor and introducing a phenomenological
cutoff for long wave-vector fluctuations they showed that the dc 
excess conductivity in YBCO films agreed well with
the modified AL expression, without the need for MT terms (up to 
$T=1.1\;T_c$). Most important, from this kind of
analysis no 3D-2D crossover was found in YBCO, in contrast with the 
ordinary LD-like crossover. \cite{hopf}\\
These interpretations are based on theoretical results obtained in 
the Gaussian approximation. Approaching
the critical region close to $T_c$ this treatment must be extended to 
take into account interactions between
fluctuations. The amplitude of the critical region is theoretically 
predicted to be experimentally accessible for these
superconductors, \cite{kap} but the actual value of the crossover 
temperature from Gaussian to critical behavior is
still debated. \cite{gauzzi,fisher}\\
Up to now, the analysis of the dynamical properties near $T_{c}$
have been performed mainly through
the comparison of experimental data with appropriate power laws of 
the reduced temperature. However,
experimental data give controversial results. For example, the dc 
conductivity above $T_c$ \cite{menego} and the
penetration depth $\lambda (T)$ below $T_c$  \cite{kamal} have been 
found to follow a 3D XY-like power law, but
Gaussian results for $\lambda (T)$ have been recently reported.
\cite{andreone} Moreover, a possible crossover
from critical to 2D Gaussian fluctuations has been reported in the 
microwave conductivity. \cite{anlage} In fact,
simple power laws or scaling behaviors, without explicit expressions 
for the various quantities, do not allow a
quantitative and complete comparison between experiments and 
theories.\\
Explicit expressions for the finite-frequency conductivity have been 
recently calculated by Dorsey \cite{d}
beyond the Gaussian approximation, using a Hartree approach; in 
this treatment, confined to an isotropic,
three dimensional superconductor, a renormalized expression for the 
3D isotropic fluctuational conductivity is
deduced, explicitly as well as in a scaling form. The basic scaling 
parameter is the temperature-dependent
correlation time $\tau \sim \xi^z$, with the correlation length 
diverging at $T=T_c$  as $\xi \sim \epsilon^{-\nu}$,
where sufficiently close to $T_{c}$, $\epsilon = (T/T_c-1)$. Measurements of the complex 
conductivity as a function of frequency \cite{booth} analyzed in
terms of the abovementioned theory have revealed a 
somehow puzzling behavior: in fact, the complex
conductivity $\sigma\left(\omega\right)$ does exhibit a scaling 
behavior close to the expected one, but the
so-obtained critical exponents, $\nu \simeq 1.2$ and $z \simeq 2.6$, are 
quite different with respect to the Gaussian
values, $\nu = 0.5$ and $z = 2$; the critical exponent $\nu$ is also 
in conflict with the prediction for the 3D XY
uncharged fluid, $\nu =2/3$. \cite{lobb} The determination of the 
critical exponents close to $T_c$  is uncertain: in
fact, a different scaling analysis of measurements of the 
frequency-dependent conductivity up to 2 GHz in zero
magnetic field gave large exponents, $\nu \simeq 1.7$ and $z \simeq 
5.6$, \cite{naki} in contrast with those previously
obtained.\\
A noticeable fact in the existing body of experimental data and 
theoretical models is that the commonly
performed analyses do not explicitly include the anisotropy, so that 
an intrinsic feature of HTSC's is lost. In particular,
while there are Gaussian theories for the anisotropic fluctuational 
conductivity, \cite{klemm} the inclusion of at
least a mass tensor in a renormalized theory for the finite-frequency 
fluctuational conductivity is still missing, at
least to our knowledge. Since in materials such as HTSC's the intrinsic 
anisotropy is one ingredient that possibly
makes the departure from Gaussian behavior experimentally observable, we think that a 
quantitative analysis of the data must be
based on the explicit inclusion of the anisotropy in the 
calculations.\\
In this paper we extend the renormalized-fluctuations theory 
developed by Dorsey \cite{d} by introducing an
anisotropic mass tensor, and we compare the results to resistive 
transitions in zero field obtained in dc and at high
frequency, above the critical temperature. In Section \ref{th}, we 
calculate the Gaussian and
renormalized-fluctuations-induced excess conductivity above $T_c$ in 
a uniaxial superconductor, subjected to an
alternating electric field along the $(a,b)$ planes, stressing the 
main differences that come out by the introduction
of the anisotropy. We write down the explicit expressions for the 
dynamical conductivity as a function of the
frequency and temperature, in terms of physical parameters (coherence 
lengths, penetration depth). In Section
\ref{ex} we briefly describe the samples under study, we sketch the 
experimental apparatures and we present the
resistive transitions in dc, at 24 and 48 GHz. In Section \ref{conc} 
we show that very good fits to the data are obtained
with the extended theory here developed, with very reasonable 
parameters. A smooth departure from Gaussian fluctuations is obtained.

\section{Theory}
\label{th}
To take into account the intrinsic anisotropy of HTSC's , we start from 
the standard Ginzburg-Landau functional
for a uniaxial anisotropic superconductor in presence of an external 
potential vector ${\bf A}$ (throughout the
paper we use Sist\`{e}me International units):

\uprule
\begin{equation}
F=\int d{\bf r}\left[ \sum\limits_{j=x,y,z}\frac{\hbar ^2}{2m_j}\left|
\left( \frac \partial {\partial r_j}-\frac{ie^{\star}}\hbar A_j\right) \psi 
\left( 
{\bf r}\right) \right| ^2+\alpha \,\left| \psi \left( {\bf r}\right) \right|
^2+\frac 12\beta \,\left| \psi \left( {\bf r}\right) \right| 
^4\right] 
\label{th1} 
\end{equation}
\downrule

where $e^{\star}=2e$ is twice the electronic charge, $m_{x,y}=m_{ab}$ and 
$m_{z}=m_{c}$ are the masses of the pair along the
main crystallographic directions, the coefficient $\alpha$ is a 
linear function of the reduced temperature $\alpha = a
\epsilon$, and $\epsilon = \ln(T/T_c)$ is the reduced temperature. 
\cite{gorkov}
Our aim is to calculate the dynamical conductivity along the $(a,b)$ 
planes (e.g. the $x$ axis) in presence of an
electric field ${\bf E}$, or equivalently, a vector potential ${\bf 
A}$. In order to calculate a dynamical property of the
system, such as the conductivity, we consider the temporal evolution 
as determined by the time dependent
Ginzburg-Landau equation

\uprule
\begin{eqnarray}
\frac 1{\Gamma _0}\left( \frac \partial {\partial t}+\frac{ie^{\star}}\hbar 
\phi
\left( {\bf r},t\right) \right) \psi ({\bf r},t)=-\frac{\delta F[\psi 
]}{
\delta \psi ^{\star}}+\zeta ({\bf r},t)=
\nonumber 
\\ 
=-\left[ \alpha +\beta \left| \psi
\right| ^2-\frac{\hbar ^2}{2m_{ab}}\left( \frac \partial {\partial 
x}-\frac{
ie^{\star}¥}\hbar A_x\right) ^2-\frac{\hbar ^2}{2m_{ab}}\frac{\partial 
^2}{\partial
y^2}-\frac{\hbar ^2}{2m_c}\frac{\partial ^2}{\partial z^2}\right] \psi
\left( {\bf r},t\right) +\zeta ({\bf r},t)
\label{th2} 
\end{eqnarray}
\downrule

where  $\Gamma_0$ is a constant relaxation time; thermal fluctuations 
are represented by the noise term $\zeta
\left( {\bf r}, t\right)$ with $\delta$ function correlation 
$\left\langle \zeta^{\star}\left( {\bf r},t\right) \;\zeta \left( {\bf 
r}^{\prime
},t^{\prime }\right) \right\rangle =\left( 2k_BT/\Gamma _0\right) 
\delta
\left( {\bf r}-{\bf r}^{\prime }\right) \;\delta \left( t-t^{\prime 
}\right) $. We choose the gauge where the scalar potential $\phi
\left( {\bf r}, t\right)=0$.\\
The calculation scheme is as follows: we first compute the 
conductivity above $T_c$ in the linear response in
the Gaussian approximation. The result will depend on the temperature 
through $\alpha (T) $. Then we renormalize
the $\alpha$ parameter by using an Hartree approximation for the 
quartic term in the GL functional. Inserting the
latter in the Gaussian conductivity we get the renormalized result. 
This approach has been used in Ref.\protect\ \onlinecite{d} to
calculate the linear and nonlinear excess conductivity in an 
isotropic superconductor. While our analysis explicitly
include the anisotropy in the calculation of the linear conductivity, 
nonlinear effects are beyond the purposes of
this paper.
The response of the system to the in-plane field ${\bf A}(t)$ is 
determined by the current operator averaged
with respect to the noise (here represented by the brackets): it can 
be expressed as a function of the correlation
function of the order parameter $C\left( {\bf r},t;{\bf r}^{\prime 
},t^{\prime }\right) =\left\langle \psi
\left( {\bf r},t\right) \;\psi ^{\star}\left( {\bf r}^{\prime },t^{\prime
}\right) \right\rangle $:

\begin{equation}
\left\langle J_x\left( t\right) \right\rangle =-
\frac{\hbar e^{\star}}{m_{ab}}\int 
\frac{d^3{\bf q}}{\left( 2\pi \right) ^3}q_xC\left[ {\bf k}={\bf 
q}-
\frac {e^{\star}} {\hbar} {\bf A}\left( t\right) ;t,t\right] \
\label{th3} 
\end{equation}

where the momentum dependence has been shifted from ${\bf k}$ to the 
new vector ${\bf q}={\bf k}+\left( e^{\star}/\hbar \right) {\bf A}\left( 
t\right) $.
As a first step (Gaussian approximation), we neglect the non linear 
term $\beta \left| \psi \right| ^2 \psi$.
Eq.\ref{th1} is then exactly solvable, and for the correlation 
function one gets:

\uprule
\begin{eqnarray}
C\left( {\bf q};t,t\right) =
2k_BT\Gamma _0
\int\limits_{0}^{+\infty }
exp
\left\{
-2\Gamma _0\alpha s-\Gamma _0\hbar ^2s
\left( 
\frac{q_y^2}{m_{ab}}
+\frac{q_z^2}{m_c}
\right)
+
\right.
\nonumber
\\
\left.
-\frac{\Gamma _0\hbar ^2s}{m_{ab}}
\left[q_x+\frac{e^{\star}}{\hbar s}\int\limits_0^sdu\left[ A_x\left( t-u\right) -A_x\left( t\right) \right] \right] ^2
\right.
\left.
-\frac{\Gamma _0e^{\star^2}}{m_{ab}} 
\left[
\int\limits_0^sdu\left(A_x\left( t-u\right) \right)^2
-\frac{1}{s}
\left(\int\limits_0^sduA_x\left( t-u\right) \right)^2
\right]
\right\}ds
\label{th4} 
\end{eqnarray}
\downrule

In the frame of the linear response, the quadratic terms in the 
vector potential can be neglected. Using the
expression found for the correlation function, Eq.\ref{th4}, the 
current operator in Eq.\ref{th3} becomes

\uprule
\begin{equation}
\left\langle J_{x} \left( t\right) \right\rangle =\frac{e^{2} }{\hbar 
^{3} }
k_{B} T\;\left( \frac{m_{c} }{\Gamma _{0} \pi ^{3} } \right) 
^{1/2}
\;\int\limits_{0}^{+\infty }ds\left[ \frac{e^{-2\Gamma _{0} \alpha s} 
}{
s^{5/2 } } \int\limits_{0}^{s}du\left[ A_{x} \left( t-u\right) -A_{x} 
\left(
t\right) \right] \right]
\label{th5} 
\end{equation}
\downrule

After Fourier transformation, one has $J_x=\left[ \sigma^\prime ( 
\omega ) +i\sigma^{\prime\prime} ( \omega)\right]\;E_{x}¥$, with

\begin{equation}
\sigma ^{\prime }(\omega )=\frac{e^2}{\hbar ^3}k_BT\;\left( 
\frac{m_c}{
\Gamma _0\pi ^3}\right) ^{1/2}\;\int\limits_0^{+\infty 
}\frac{ds}{s^{5/2}}
e^{-2\Gamma _0\alpha s}\frac{1-\cos \,\omega s}{\omega ^2}
\label{th6} 
\end{equation}

\begin{equation}
\sigma ^{\prime \prime }(\omega )=\frac{e^2}{\hbar ^3}k_BT\;\left( 
\frac{m_c
}{\Gamma _0\pi ^3}\right) ^{1/2}\;\int\limits_0^{+\infty 
}\frac{ds}{s^{5/2}}
e^{-2\Gamma _0\alpha s}\frac{\omega s-\sin \,\omega s}{\omega ^2}
\label{th7} 
\end{equation}

After integration, Eqs.\ref{th6} and \ref{th7} can be written as:

\begin{eqnarray}
\sigma _g(\omega )=\sigma ^{\prime }(\omega )+i\sigma ^{\prime \prime
}(\omega )=
\nonumber
\\
=\frac{e^2}{32\hbar \,\xi _{c0}^{}}\frac 1{\epsilon 
^{1/2}}\left[
S_{+}\left( \frac \omega \Omega \right) +i\,S_{-}\left( \frac \omega 
\Omega
\right) \right]
\label{th8} 
\end{eqnarray}

where $S_+ (x)$ and $S_- (x)$ are the scaling functions
as can be found in Ref.\protect\ \onlinecite{d} (the subscript $ g $ means that this 
result is obtained in the Gaussian approximation), and
they have the property that $S_{+} (x\to 0)=1$ and $S_{-} (x\to 
0)=0$. The characteristic frequency $\Omega$ is

\begin{equation}
\Omega=2\Gamma_0\alpha= \frac {32 k_B T}{h}\epsilon
\label{th9} 
\end{equation}

where the relaxation time $\Gamma_0=\left(8k_BT / \hbar \pi a 
\right)$ is evaluated from the microscopic theory \cite{skoc} 
and $\xi_{c0}= {\hbar}/{\left( 2 m_{c} a \right)^{1/2}}$ is 
the zero temperature c-axis correlation length.
Before proceeding further to the renormalization, some comments are 
in order. Eq.\ref{th8} contains all the
previously obtained results in various limits: as $\omega \to 0$ 
Eq.\ref{th8} gives the dc, anisotropic AL
result. \cite{hopf,mt} As expected, at nonzero frequencies the 
conductivity does not diverge at $T_c$, due to the
vanishing of the scaling functions $S_{\pm}¥$ when written in terms of the 
temperature. Moreover, the result  in Eq.\ref{th8}
agrees with the one calculated by a different approach by Klemm.
\cite{klemm} The isotropic result is recovered
by simply putting $\xi_{c0}=\xi$. The introduction of the anisotropy 
leads to an enhancement by a factor of
$\gamma= \xi_{ab0}/\xi_{c0}$ in the prefactor of the fluctuation 
conductivity, as seen by the fact that Eq.\ref{th8}
contains only the short coherence length $\xi_{c0}$. However, {\em in 
the Gaussian approximation}, the
characteristic frequency remains unchanged with respect to the 
isotropic result. This is no longer true in the
renormalized regime, as we show below.\\
Approaching  $T_c$, the Gaussian 
approximation breaks down. We extend our
calculation to this region by considering the effects of the 
interaction term $\beta \left| \psi \right|^2\psi$ of the
Landau-Ginzburg functional through the Hartree approximation: we 
replace the non-linear term by its
average value and put it in a renormalization of the parameter 
$\alpha$. The renormalized parameter $\tilde{\alpha}$

\begin{equation}
\tilde{\alpha } =\alpha +\beta \left\langle \left| \psi \right| ^{2}
\right\rangle =\alpha +\beta \int \frac{d^{3} {\bf q}}{(2\pi )^{3} } 
\; C\left(
{\bf q};t,t\right)
\label{th10} 
\end{equation}

represents the renormalized temperature dependence and it is defined 
through this self-consistency equation.\\
All the quantities above calculated in the Gaussian approximation 
contain the temperature dependence
through the parameter $\alpha$; hence, the relations found can be 
easily extended to the critical region by
replacing $\alpha \rightarrow \tilde{\alpha}$. In particular, the 
correlation function is formally identical to the one
determined by means of Eq.\ref{th4}; evaluating it with an electric 
field along the $x$-axis in the frame of the linear
response approximation, Eq.\ref{th10} becomes:

\begin{equation}
2m_{ab}\tilde{\alpha}=2m_{ab}\left( \alpha -\alpha _c\right) 
-\frac{\beta
k_BT}{\hbar ^3\pi }\left( m_{ab}^3\;m_c\right) ^{1/2}\left( 
2m_{ab}\tilde{
\alpha}\right) ^{1/2}
\label{th11} 
\end{equation}

where $\alpha_c$ is the bare $\alpha$ parameter evaluated at the 
renormalized critical temperature $\tilde{T_c}$ at
which the parameter $\tilde{\alpha}$ vanishes: $\alpha _c=\alpha 
(T=\tilde{T}_c)=a\;\ln \left( \tilde{T}_c / T_c \right)$.\\
	The self consistency equation, Eq.\ref{th11}, can be usefully interpreted 
as the relation which determines the renormalized correlation length 
along the $(a,b)$ planes 
$\tilde{\xi_{ab}}=\left(\hbar^2/2m_{ab}\tilde{\alpha}\right)^{1/2}$:

\begin{equation}
\tilde{\xi}_{ab}(T)=
\frac {w\;\kappa ^2\;\gamma \;\xi _{ab0}^2}{\ln \left(T/\tilde{T}_c \right) } 
\left[
1+
\left(
1+\frac{\ln \left( T / \tilde{T}_c \right)}
{w^2\;\kappa ^4\;\gamma^2\;\xi _{ab0}^2}
\right)^{1/2}
\right]
\label{th12} 
\end{equation}

where $w=\left( e^{2} \,\mu _{0} \,k_{B} T/\pi \,\hbar ^{2} \right) 
$, $\mu_{0}$ is the magnetic permeability of vacuum, $\gamma = \left( \xi_{ab0} / \xi_{c0} \right)$ is the anisotropy 
factor and $\kappa = \left( \lambda_{ab0} /
\xi_{ab0} \right) $ is the Ginzburg parameter, which is related to 
the coefficient $\beta$ of the Landau-Ginzburg
functional through the London equation. \cite{bula} The renormalized 
coherence length follows the usual, Gaussian
temperature behavior $\epsilon^{-\nu}$ with $\nu = 1/2$ sufficiently 
far away from $T_c$, but approaching the
critical temperature it diverges with the critical exponent $\nu = 
1$, as depicted in Fig.\ref{figcsi}. A smooth
crossover between these two power laws is then obtained by varying 
the temperature (Fig.\ref{figcsi}). It is
interesting to note that, while the exponent $\nu = 1$ is recovered 
only very close to $T_c$, a substantial departure
from the Gaussian value $\nu = 1/2$ is obtained at rather high 
temperatures, and this regime is very well
approximated by an exponent $\nu = 2/3$.
Once the renormalized coherence length is obtained, the conductivity 
can be immediately written down by
substituting $\xi_{ab}(T)$ with $\tilde{\xi}_{ab}(T)$, and one gets

\uprule
\begin{equation}
\tilde{\sigma}\left( \omega ,T\right) =\frac{e^2}{32\hbar }\gamma 
\frac{
\tilde{\xi}_{ab}\left( T\right) }{\xi _{ab0}^2}\;\left[ S_{+}\left( 
\frac
\omega {\tilde{\Omega}\left( T\right) }\right) +iS_{-}\left( \frac 
\omega {
\tilde{\Omega}\left( T\right) }\right) \right] 
=\tilde{\sigma}_{dc}\left(
T\right) \,\left[ S_{+}+iS_{-}\right]
\label{th13} 
\end{equation}
\downrule

where 

\begin{equation}
\tilde{\Omega } \left( T\right) =\frac{32\,k_{B} \,T}{h} 
\;\left( 
\frac{\xi _{ab0} }{\tilde{\xi } _{ab} \left( T\right) } \right) ^{2}
\label{th14} 
\end{equation}

and $\tilde{\tau}=1/\tilde{\Omega}(T)$ plays the role of the 
renormalized scattering time. Expression (\ref{th13})
takes then the form of the renormalized dc excess conductivity 
$\tilde{\sigma}_{dc}(T)$  times a
frequency-dependent contribution. As it can be noted from 
Eq.\ref{th12}, the expression for the fluctuation
conductivity depends on a limited number of parameters, namely the 
bare quantities $\xi_{ab0}$, $\gamma$ and
$\kappa$, being $\tilde{T}_c$ the only renormalized parameter. We note that now the
(renormalized) scattering time {\em does } depend on the anisotropy, 
differently from the Gaussian result,
Eq.\ref{th9}.
As expected, in the limit $\gamma=1$ our results coincide with the 
isotropic calculation. \cite{d} We stress,
however, that the anisotropy $\gamma$ does not enter in a trivial way 
in the renormalized quantities: it enters
through different combinations in the prefactor of $\tilde{\sigma}$ 
and in the renormalized scattering time. In
particular, our calculation cannot be simply mapped onto the 
isotropic result by the use of some "lumped"
parameter in the fitting: $\gamma$ is an {\em independent} parameter.

\begin{figure}
\centerline{\psfig{figure=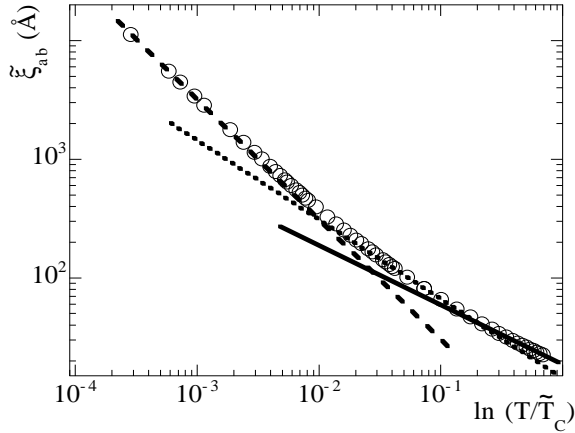,height=6cm,width=8cm,clip=,angle=0.}}

\caption{Renormalized coherence length as a function of the 
temperature, $\tilde{\xi}_{ab} $ (open circles).
Parameters appropriate to the fitting of the data on sample IV were 
used. Various power laws are depicted, in the
form $\tilde{\xi}_{ab} \sim \left[\left( ln(T/\tilde{T}_c 
)\right)\right]^{-\nu}$: 
Gaussian ($\nu = 1/2$, full line), critical ($\nu = 1$, dashed
line) and $\nu = 2/3$ (dotted line).}
\label{figcsi}
\end{figure}

\section{Experimental Section}
\label{ex}

Measurements of the microwave and dc resistivity in nominally zero 
field were performed. Five YBCO thin
films, grown by different methods \cite{boffa,sparvieri,merlo} were 
investigated. All samples were highly c-axis
oriented, as indicated from the $\theta-2\theta$ rocking curve. 
Twinning is largely present in all samples (as usual
in films). Thicknesses ranged from 0.08 to 0.5 $\mu$m. The main features 
are presented in Table \ref{t1}. It is worth
mentioning that samples II and III are of inferior quality, as can be 
seen, e.g.,by the fact that the normal state
resistivity is 2-3 times higher than in the other films.\\
Microwave resistivity measurements were performed on as-deposited 
samples I-IV. The microwave response
was investigated at 48 and 24 GHz, in samples I-III and IV, 
respectively. Extensive descriptions of the experimental
systems have been given previously, \cite{fastampa,silva} and we give 
here only a short sketch.
Two experimental systems were employed. In both cases we made use of 
the cavity-end-wall-replacement
method: the sample is mounted in order to replace one end-wall of a 
mechanically tunable right-cilynder resonant
cavity. The cavities were designed to work in absorption in the 
TE$_{011}$ mode, at 24 [cavity (a)] and 48.2 GHz [cavity
(b)], with quality factors of about 15000 and 6000, respectively. The 
relatively low quality factor prevented an
accurate measurement of the absolute surface resistance below $\sim 
70 K$, but allowed us to obtain reliable
measurements in the whole transition range and well above $T_c$. As 
described in Ref.\protect\ \onlinecite{silva}, the unloaded
quality factor $Q$ of the cavities was measured by recording the 
(Lorentzian) resonance shape as a function of the
slowly ($\sim0.2$ K/min) increasing temperature. Changes of $Q$ 
reflect the changes in the microwave surface
resistance. \cite{silva} A calibration of the cavity response is in 
principle needed to obtain the absolute surface
resistance, but since below $\sim$ 70 K the changes in $Q$ cannot be 
resolved due to the reduced sensitivity of the
cavities, data presented as:

\begin{equation}
R_{S} (T)-R_{S} (70K)=G\,\left[ \frac{1}{Q(T)} -\frac{1}{Q(70K)} 
\right]
\label{ex1} 
\end{equation}

are independent on the calibration (a detailed discussion can be 
found in Ref.\protect\ \onlinecite{silva}). This is equivalent to take as
zero the low-temperature value of $R_S$. Here $G$ is a known 
geometrical factor.\\
The real part of the microwave resistivity is directly obtained from 
the data due to the reduced sample
thickness $d$. For samples thinner than twice the penetration depth 
(the skin depth $\delta$ in the normal state,
the London penetration depth $\lambda$ in the superconducting state) 
the measured surface resistance directly
gives the real part of the resistivity through:

\begin{equation}
Re \left[ \rho (T)\right] \cong R_S(T)\,d
\label{ex2} 
\end{equation}

A detailed study of the applicability range of this approximation can 
be found in Ref.\protect\ \onlinecite{silva2}. We notice, however,
that we are interested in measurements close and above $T_c$, where 
$\lambda$ is much longer than the
zero-temperature value. Consequently, Eq.\ref{ex2} is valid up to (at 
least) 3\%,\cite{silva2} also in the thicker film.
\uprule
\begin{table}
\begin{tabular}{ccccccccc}
Sample & Thickness($\mu$m) & Substrate & $\omega$(GHz) & $T_i$ & 
$\tilde{T}_c$ & $\xi_{ab0}$(\AA)& $\lambda_{ab0}$(\AA) & $\gamma$ \\ 
I & 0.3 & LaAlO$_3$ & 48.2 &  86.2 & 86.7 & 14.4 & 1300 & 6.4\\ 
II&  0.08 & LaAlO$_3$ & 48.2 & 86.5  & 86.8 & 17.5 & 1300 & 5\\ 
III&  0.5 & LaAlO$_3$ & 48.2 & 87.2 & 87.2 & 16.5 & 1500 &5.4 \\ 
IV& 0.12 & LaAlO$_3$ & 24.0 & 84.6 & 84.4 &  14.3 & 1500 &6.6\\ 
V&  0.1 & SrTiO$_3$ & dc & 89.7 & 89.6 & 15.0 & 1100 & 6.3
\end{tabular}
\caption{Sample characteristics, measuring frequency ($\omega$), 
measured inflection temperature 
($T_{i}$) and fit parameters ($\tilde{T}_c$, $\xi_{ab0}$, $\lambda_{ab0}$,$\gamma$).}
\label{t1}
\end{table}
\downrule

The data for the so-obtained microwave resistivity on four samples 
as a function of the temperature are
reported in Figs.\ref{figsp} and \ref{figlas} in terms of 
$\Delta\rho\left(T\right)=\rho\left(T\right)-\rho\left(70 K\right)$, 
accordingly to Eqs.\ref{ex1}  and \ref{ex2} . 
The behavior is essentially linear at high temperatures, and gradually
bends approaching the transition. The inflection temperatures of the 
microwave resistivity, $T_{i}$, in the various samples are
reported in Table \ref{t1}. Sample IV was measured at 24 GHz, and 
samples I-III at 48.2 GHz.
Additionally, we performed dc resistivity measurements on a 
patterned sample. Sample V was patterned in a 2
mm long, 100 $\mu$m wide strip by ion milling. The dc resistivity was 
measured through a four-probe lock-in method,
with the current oscillating at 20 Hz. Low current density ($\sim$10 
A/cm$^2$) was used. Data were collected upon cooling
and warming in a commercial helium flow cryostat. No differences were 
observed in the corresponding resistive
transition, within the voltage ($\sim$ 5 nV) and temperature ($\sim$ 
0.005 K) sensistivities. Zero resistance (within our
resolution) was attained at $T_z$=88.8 K.

\section{Discussion and conclusions}
\label{conc}

We proceed here to the fitting of the zero-field resistive 
transitions above $T_c$ on the five samples with the
proposed model. First of all, we write the total conductivity as the 
sum of the normal and fluctuational terms, so
that the total resistivity is given by

\begin{equation}
\rho (T,\omega )=\frac{1}{\sigma _{n} (T)+\tilde{\sigma } _{fl} 
(T,\omega )}
=\frac{1}{\frac{1}{\rho _{n} (T)} +\tilde{\sigma } _{fl} (T,\omega )}
\label{conc1} 
\end{equation}

where we have assumed that the normal state relaxation time is much 
smaller than $1/\omega$. Here, $\tilde{\sigma}_{fl}(T,\omega)$ is the fluctuational 
conductivity and $\rho_n(T)$ is the measured normal-state 
resistivity. The latter is linear above $\sim$120 K, and it was
linearly extrapolated down to $T_{c}$.\\
Fits of the experimental data with the real part of Eq.\ref{conc1} 
can now be made. The theoretical expression
depends on four independent parameters, that can be chosen to be the 
bare $\xi_{ab0}$, $\lambda_{ab0}$ and $\gamma$, and the
renormalized critical temperature $\tilde{T}_c$ (which is, in this 
frame, the experimentally observable critical
temperature). With the present
interpretation, no information comes from the data for the bare 
critical temperature, $T_c$.\\
Before commenting on the fits, a few notes should be added. First of 
all, as common to many calculations of the
fluctuational conductivity,\cite{ike} the theory overestimates the 
fluctuational contribution for $T \gg T_c$. This is true
also at zero frequency. In fact, calculations of the dc excess 
conductivity in the Gaussian approximation, explicitly
including a high-${\bf q}$ cutoff in the various integrations in the 
momentum space show a drastic and sharp
suppression of $\tilde{\sigma} \left( T, \omega = 0 \right)$ at 
sufficiently high $T$ (above $\sim$120 K for YBCO).
\cite{hopf,gauzzi,cimberle} 
\begin{figure}
\centerline{\psfig{figure=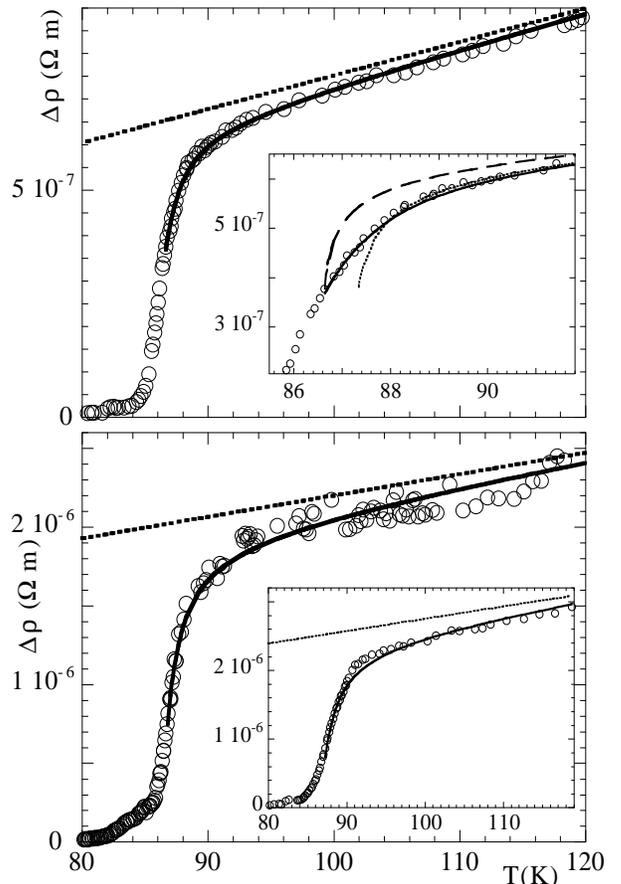,height=12cm,width=8cm,clip=,angle=0.}}

\caption{Measurements of the microwave resistivity at 48.2 GHz on 
sputtered samples I-III (open circles) and fits
through Eq.\ref{conc1} (continuous lines). Normal state 
resistivity: dotted lines. Upper panel: data for the
optimized sample (sample I), with lower normal state resistivity. In 
the inset: enlargement in a restricted
temperature range. The anisotropic Gaussian fits through Eq.\ref{th8} 
are also reported for comparison: dashed line (same parameters as for 
the renormalized fit) and dotted line ($T_{c}$=87.3 K, 
$\xi_{c}=1.7\AA$). The Gaussian fits do not reproduce the shape of 
the transition. Lower panel: data for lower quality samples II 
(main panel) and III (inset). Scattering of the
data above $T_c$ are due to the sensitivity limit of the cavity.}
\label{figsp}
\end{figure}
Inclusion of such a cutoff in the 
frequency-dependent fluctuational conductivity is
beyond of the scope of this paper. Instead, we will 
phenomenologically take into account this effect by making the
physically reasonable assumption that at sufficiently high 
temperature (say, 150 K) the fluctuational conductivity is
no longer resolved in the measurements. As a consequence, we will 
write the dc excess conductivity in Eq.\ref{th13} 
(and then in Eq.\ref{conc1}) as $\tilde{\sigma } _{dc} \left( 
T\right) -\tilde{\sigma } _{dc} \left(
150\,K\right) $. Other possible choices consist essentially of 
taking a normal-state resistivity
which is higher than the measured one \cite{ike} or with a different 
shape with respect to the linear extrapolation.
\cite{holm} However, our present choice keeps as much contact as 
possible with the measured data well above
$T_c$, and does not introduce additional parameters.\\
A second point comes from the fact that the HF renormalization here 
presented is perfomed in the adiabatic limit, that is $\omega /
\tilde{\Omega }(T)\ll1$. As a consequence, we expect that $\omega / 
\tilde{\Omega }(T) \approx 1$ is a limit for the accurate
applicability of such calculations, while the main features might apply also 
beyond this point until the HF approach breaks down very near $T_{c}$.
Since $\tilde{\Omega }\left( T \to \tilde{T}_c\right) \rightarrow 0$, 
 all the fits were performed self-consistently, excluding the data points 
with temperatures lower than those for which
$\omega / \tilde{\Omega}(T) = 1$. This requirement, together with 
the expected cutoff at high temperature,
resulted in fits being performed using the data from about 0.5 K 
above the critical temperature, up to $\sim$120 K.\\
With these cares, we performed the fits with the fluctuational 
conductivity here calculated. The results are
reported in Figs.\ref{figsp},\ref{figlas} and \ref{figdc}. As can be seen, all the 
fits are very satisfying, and the anisotropic
renormalized fluctuational conductivity seems to be a good 
description of the data. All the parameters obtained
are in the range of commonly reported values (see Table \ref{t1}).
\cite{ybco,krusin,harsh,felner,hao,welp,farrell} 
In particular, we note that the low-quality films (samples II and III) have lower 
$\gamma$, as expected and usually found in
literature. As a matter of fact, all the $\tilde{T}_c$'s almost 
coincide with the inflection of the transition, well below
the first onset of the superconductivity. If one allows the bare 
critical temperature to be at the onset of
superconductivity, one has a rough estimate $\left| \tilde{T}_c-T_c 
\right| \sim $3K. 
While the condition $\omega / \tilde{\Omega
}(T) = 1$ is reached about $\sim$ 0.5 K above $\tilde{T}_c$ (slightly 
depending on the sample), we note that the fits are in fairly good
agreement with the data even down to $\tilde{T}_c$. It should be mentioned 
that preliminary analysis of the flux-flow
resistivity \cite{silva3} at 48ÊGHz below $T_c$ on sample I gave an 
independent estimate for the coherence length, in agreement with the value 
here obtained from zero-field fluctuational conductivity {\em
above} $T_c$.\\%
\begin{figure}
\centerline{\psfig{figure=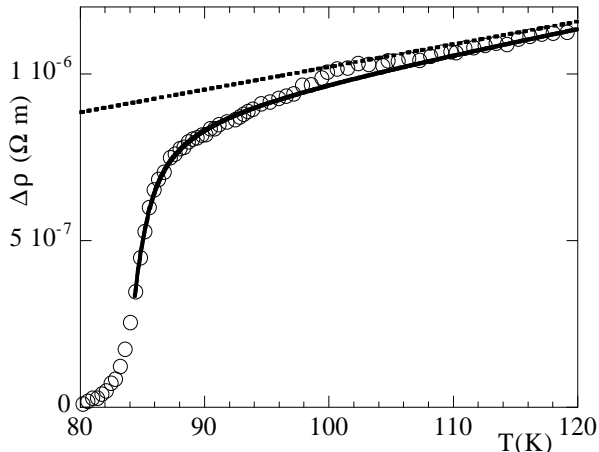,height=6cm,width=8cm,clip=,angle=0.}}
\caption{As in Fig.2, for the microwave resistivity at 24 GHz on the 
laser-ablated sample IV.}
\label{figlas}
\end{figure}
We mention that our data might be equally well fitted with the 
original isotropic theory \cite{d}. However in this case the 
parameters attain unrealistic values. As an example, on sample I one 
would get a fit with the isotropic theory almost indistinguishable 
from the one obtained with the anisotropic theory with 
$\xi=2.6\AA$. 
Since here $\xi$ is an isotropic coherence 
length, it is reasonable to assume 
$\xi=(\xi_{ab}^{2}\xi_{c})^{1/3}=\xi_{ab}/\gamma^{1/3}$. For 
$\xi_{ab}\approx15\AA$ \cite{} one would get $\gamma\approx 200$, 
more than an order of magnitude higher than known values in YBCO.
The Gaussian anisotropic theory (Eq.\ref{th8}) can be made to fit the 
data only in the high part of the resistive transitions. An 
example is reported in the inset of Fig.\ref{figsp}. It is apparent 
that the Gaussian approximation does not reproduce at all the shape 
of the resistive transition.
A final note on the 3D-2D dimensional crossover that might be 
expected at high temperatures, owing to the
decrease of the $c$-axis coherence length: being our calculation 
explicitly 3D, such a crossover is not included
there. However, it seems that at least up to $\sim$120 K the data are well 
described by the anisotropic 3D theory. This
behavior is completely analogous to the results obtained in dc,
\cite{hopf,gauzzi} where a Gaussian analysis of the
data sufficiently above $T_c$ did not show any dimensional crossover, 
the departure from the AL anisotropic 3D
behavior being fully accounted for by the introduction of the cutoff 
in ${\bf q}$-space. It would be consistent also
with the fact that, on a pure numerical ground, the renormalized 
coherence length is longer than the bare one, so
that the crossover temperature shifts to higher $T$, where this 
phenomenon might be hardly distinguishable from
the short-wavelength-fluctuation regime. This point deserves further 
study in the future.\\
\begin{figure}
\centerline{\psfig{figure=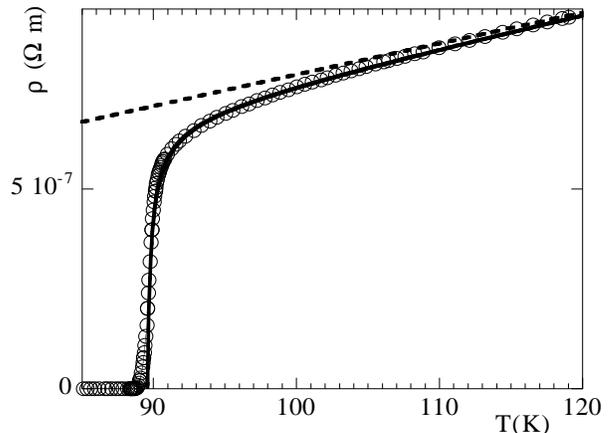,height=6cm,width=8cm,clip=,angle=0.}}
\caption{As in Fig.\ref{figsp}, for the dc resistivity on the 
laser-ablated sample V. For clarity, only 5\% of data are shown. The theory reproduces fairly well
the shape of the transition above $T_c$ also in the limit of zero 
frequency.}
\label{figdc}
\end{figure}
In conclusion, we have developed a theory for the finite-frequency 
fluctuational conductivity in an anisotropic
uniaxial superconductor beyond the Gaussian approximation. We have 
shown that the anisotropy ratio $\gamma$
enters in a non-trivial way in the expression for the renormalized 
conductivity. We have performed measurements
of the dc and microwave resistivity above $T_c$ in several YBCO films 
of different quality and preparation process. In all cases the temperature dependence of the 
resistivity at all the frequencies
investigated could be well described by the theory here developed 
from slightly above $T_c$ up to $\sim$ 120 K, with
values.
Above $\sim$ 120 K the theory does not apply, and more extensions are 
needed.

\section{acknowledgments}
We are indebted to V.Merlo at the University ``Tor Vergata'' (Rome) 
and V.Boffa at the ENEA laboratories (Frascati) 
for supplying samples IV and V, respectively.

\end{multicols}

\end{document}